\documentclass[preprint,5p]{elsarticle}
\usepackage{hyperref,amsmath,amsfonts,epstopdf,float}
\usepackage{subfig}
\begin{document}

\begin{frontmatter}

\title{On Complex Langevin Dynamics and the Evaluation of Observables}
\cortext[cor1]{Corresponding author}
\author[ad]{A. Durakovic}
\ead{amel@nbi.dk}

\author[eca]{E. C. Andre}

\author[at]{A. Tranberg\corref{cor1}}
\ead{anders.tranberg@uis.no}
\address[ad]{Niels Bohr International Academy and Discovery Center, Niels Bohr Institute, Blegdamsvej 17, DK-2100 Copenhagen, Denmark}
\address[eca]{Niels Bohr Institute, Blegdamsvej 17, DK-2100 Copenhagen, Denmark}
\address[at]{Faculty of Science and Technology, Kjell Arholms gate 41, University of Stavanger, N-4036 Stavanger, Norway}


\begin{abstract}
In stochastic quantisation, quantum mechanical expectation values are computed as averages over the time history of a stochastic process described by a Langevin equation. Complex stochastic quantisation, though theoretically not rigorously established, extends this idea to cases where the action is complex-valued by complexifying the basic degrees of freedom, all observables and allowing the stochastic process to probe the complexified configuration space. We review the method for a previously studied one-dimensional toy model, the U(1) one link model. We confirm that complex Langevin dynamics only works for a certain range of parameters, misestimating observables otherwise. A curious effect is observed where all moments of the basic stochastic variable are misestimated, although these misestimated moments may be used to construct, by a Taylor series, other observables that are reproduced correctly. This suggests a subtle but not completely resolved relationship between the original complex integration measure and the higher-dimensional probability distribution in the complexified configuration space, generated by the complex Langevin process. 
\end{abstract}

\end{frontmatter}

\section{Introduction}
In a quantum theory, the central quantities of interest are the expectation values of various physical observables $\{O\}$, and these may in a Euclidean theory be represented in terms of path integrals
\begin{align}
\langle O \rangle = \frac{1}{Z} \int \mathrm{d}\{x\} \,O(\{x\}) e^{-S_E(\{x\})}\textrm{,}
\end{align}
where $\{x\}$ is the set of basic degrees of freedom of the system, $S_E(\{x\})$ is the dimensionless ($\hbar = 1$) Euclidean action and $Z = \int \mathrm{d}\{x\} \, e^{-S_E(\{x\})}$ is the partition function.
One approach to calculating these expectation values, which is also readily implemented for numerical computation, is stochastic quantisation (see \cite{phda} for a review).
A fictitious time dimension $\theta$ is introduced with respect to which the system wanders within the space of possible configurations $\{x\}$ in accordance with a Langevin equation derived from $S_E$.
The expectation value $ \langle O \rangle$ can then be computed as the average of all values $O(x)$ assumed on a path $x(\theta)$ in the limit as $\theta \to \infty$.

In the following, we will consider a system with only one degree of freedom, $x$. In terms of the drift force $D_x = - \partial S_E/\partial x$, and a Gaussian stochastic force $\eta$, the Langevin equation reads explicitly
\begin{align}
\dot{x}(\theta) = D_x(x(\theta))+\eta(\theta) = - \frac{\partial}{\partial x} S(x)\Big|_{x\to x(\theta)} + \eta(\theta)\textrm{.}
\end{align}
The probability distribution $P(x,\theta)$ corresponding to the stochastic dynamics obeys the Fokker-Planck equation
\begin{align}
\frac{\partial}{\partial \theta} P(x,\theta) = \left(-\frac{\partial}{\partial x} D_x(x)+ D_{xx} \frac{\partial^2}{\partial x^2}\right) P(x,\theta)\textrm{.}
\end{align}
where $D_{xx}$ is the diffusion coefficient, equal to one in this case. The equilibrium probability distribution satisfies $\partial P/\partial t = 0$ and it is straightforward to check that it is satisfied by $P^\mathrm{eq}(x) = C e^{\int^x \mathrm{d}x' D_x(x')/D_{xx}}$ where $C$ is a normalisation constant. Expectation values in the equilibrium limit which is also the infinite time limit become
\begin{align}
\langle O \rangle = C \int \mathrm{d}x \, O(x) e^{\int^x \mathrm{d}x' D_x(x')/D_{xx}} = \frac{1}{Z} \int \mathrm{d}x \, O(x) e^{-S_E}
\end{align}
as desired. This result can be generalised to any number of degrees of freedom and Euclidean quantum field theories as long as they have real-valued actions~\cite{phda}.

A problem arises when the action is complex-valued. Then the factor $e^{-S}$ cannot be interpreted as the equilibrium probability distribution of a stochastic process and the strict mapping of the problem to Langevin dynamics fails. Such actions, for instance, arise in field theories containing fermions with non-zero chemical potential, for evolution in real time and certain toy models. An appealing solution to the problem is to separate the real and imaginary parts of the action and use the former as the real-valued probability distribution while including the latter in the observable of interest. This is known as reweighting, but in many interesting cases, it results in bad statistics of the observable $\langle O \rangle = \langle O e^{i S_I}\rangle_{\mathrm{Re}}/\langle e^{i S_I} \rangle_{\mathrm{Re}}$ where $\mathrm{Re}$ denotes expectation values calculated in the theory with only the real-valued action and $S_I$ is the imaginary part of that action. This phenomenon is a variant of the ``sign'' problem.

Complex Langevin dynamics has been proposed to deal with complex-valued actions~\cite{parisi,Ambjorn,aartsu1,Sexty1}. The idea is to complexify all observables with the simple substitution $x\to x+iy$ such that $O(x) \to O(x+iy)$ and define a flow in two real dimensions. The previously defined drift force is also complexified and the new flow in the two dimensions $x$ and $y$ is determined by the real and imaginary parts of this complexified drift force, respectively. Gaussian noise may be added to either or both of the $x$ and $y$ component flow equations but is usually added only to the $x$ component.

The very long trajectory that probes the space of possible configurations traces out some probability distribution $P(x,y)$, and it is then conjectured that the average of $O(x+iy)$ over this distribution matches the path integral average such that
\begin{align}
\int O(x+iy) P(x,y) \overset{?}{=} \frac{1}{Z} \int O(x) e^{S_R + iS_I} = \langle O \rangle\textrm{.} \label{eq:conj}
\end{align}
Finding $P(x,y)$ by solving the Fokker-Planck equation corresponding to this flow is very nontrivial but the numerical Langevin evolution is straightforward. Equation (\ref{eq:conj}) may then be conjectured to apply to any observable and in principle provide for an exact quantisation of the complex-valued action. This statement implies a very strong relation between the complex integration measure on the RHS, and the two-dimensional probability distribution (also an integration measure) on the LHS. We will return to this below. 

Results are promising. It seems to evade instances of the sign problem~\cite{aartsfield,Sexty1,Sexty2,Sexty3} and the procedure correctly reproduces expectation values of certain theories. However, there exists no proof of convergence and it is still unclear why, and under which circumstances this prescription works. Very promising result have recently been obtained by \cite{Seiler1,Seiler2}.

In the following, we will consider one of the successful applications of complex Langevin dynamics, the $U(1)$ one link model. In the next section, we will introduce the model and compute a set of observables, reproducing known results~\cite{Ambjorn1, Ambjorn2,,aartsu1}. In Section~\ref{sec:limits}, we will then test the limits of the agreement and point out some observables which do not converge correctly, except in certain combinations. We conclude in Section~\ref{sec:concl}. 

\section{The U(1) one link model and successful complex Langevin dynamics}
The $U(1)$ one link model is defined in terms of a single degree of freedom $x$ and has the partition function 
\begin{align}
Z &= \frac{1}{2\pi} \int_{-\pi}^{\pi} \mathrm{d}x \, e^{-\beta \cos(x)} (1+\kappa \cos(x-i \mu)) \nonumber \\ &= \frac{1}{2\pi} \int_{-\pi}^{-\pi} \mathrm{d}x \, e^{-S_\mathrm{eff}}
\end{align}
and therefore we have $S_\mathrm{eff} = -\beta \cos(x) - \log (1+\kappa \cos(x-i \mu))$. The model has three parameters: $\beta$, which appears like an inverse temperature; $\kappa$, which appears like a coupling or ``hopping'' parameter; and $\mu$, which appears like a chemical potential.
The toy model emulates a lattice gauge theory of a single Abelian link variable $e^{ix}$ with a fermion determinant $\det(M) = \allowbreak1+\kappa \cos(x-i \mu)$ that is complex-valued due to the presence of a chemical potential.
Expectation values are defined by
\begin{align}
\langle O \rangle = \frac{1}{Z} \frac{1}{2\pi} \int \mathrm{d}x\, O(x) e^{-S_{\mathrm{eff}}}\textrm{.}
\end{align}
In \cite{aartsu1}, four observables mimicking the Polyakov loop $\langle U \rangle = \langle e^{ix} \rangle$, the conjugate Polyakov loop $\langle U^{-1} \rangle = \langle e^{-ix} \rangle$, the plaquette $\langle \cos(x) \rangle$ and the density $\langle n \rangle = \partial \log Z /\partial \mu$ were considered. These observables can be calculated analytically and a direct comparison made with the numerical estimates. The phase of the determinant $\langle e^{2i\phi} \rangle $, an indicator of the severity of the sign problem, was also considered though we will not discuss it here.

After complexification of the variable $x \to x+iy$, and discretisation of the fictitious time into steps of size $\epsilon$ such that $\theta = n \epsilon$ where $n$ is integer, the Langevin equations reduce to
\begin{align}
x_{n+1} &= x_{n} + \epsilon K_x(x_n,y_n) + \sqrt{\epsilon} \eta_n\textrm{,} \\
y_{n+1} &= y_{n} + \epsilon K_y(x_n,y_n) \textrm{,}
\end{align}
with
\begin{align}
K_x = -\sin(x)\left(\beta \cosh(y) + \kappa \frac{\cosh(y-\mu)+\kappa \cos(x)}{D(x)}\right)\textrm{,}
\end{align}
and
\begin{align}
K_y = - \kappa \sinh(y-\mu) \frac{\cos(x) + \kappa \cosh(y-\mu)}{D(x)} \nonumber \\ -\beta \cos(x) \sinh(y)\textrm{,}
\end{align}
where $D(x) = (1+\kappa \cos(x) \cosh(y-\mu))^2 + (\kappa \sin(x) \sinh(y-\mu))^2$. It was here used that $K_x = -\mathrm{Re}\, \partial S_\mathrm{eff}/\partial x |_{x\to x+iy}$ and $K_y = - \mathrm{Im}\,\partial S_\mathrm{eff}/\partial x |_{x\to x+iy}$. The noise satisfies $\langle \eta_n \rangle = 0$ and $\langle \eta_n \eta_{n'} \rangle = 2 \delta_{n n'}$ (zero mean and variance equal to $2$). From some random initial condition $(x_0,y_0)$, we can simply evaluate these equations in steps to produce a trajectory in the complex configuration space. The estimate of the observable then becomes $\langle O \rangle = 1/N \sum_{k=0}^N O(x_k+iy_k)$.

\begin{figure}[H]
\centering
\subfloat[][The Polyakov loop $\langle e^{ix} \rangle$]{
	\includegraphics[scale=0.23]{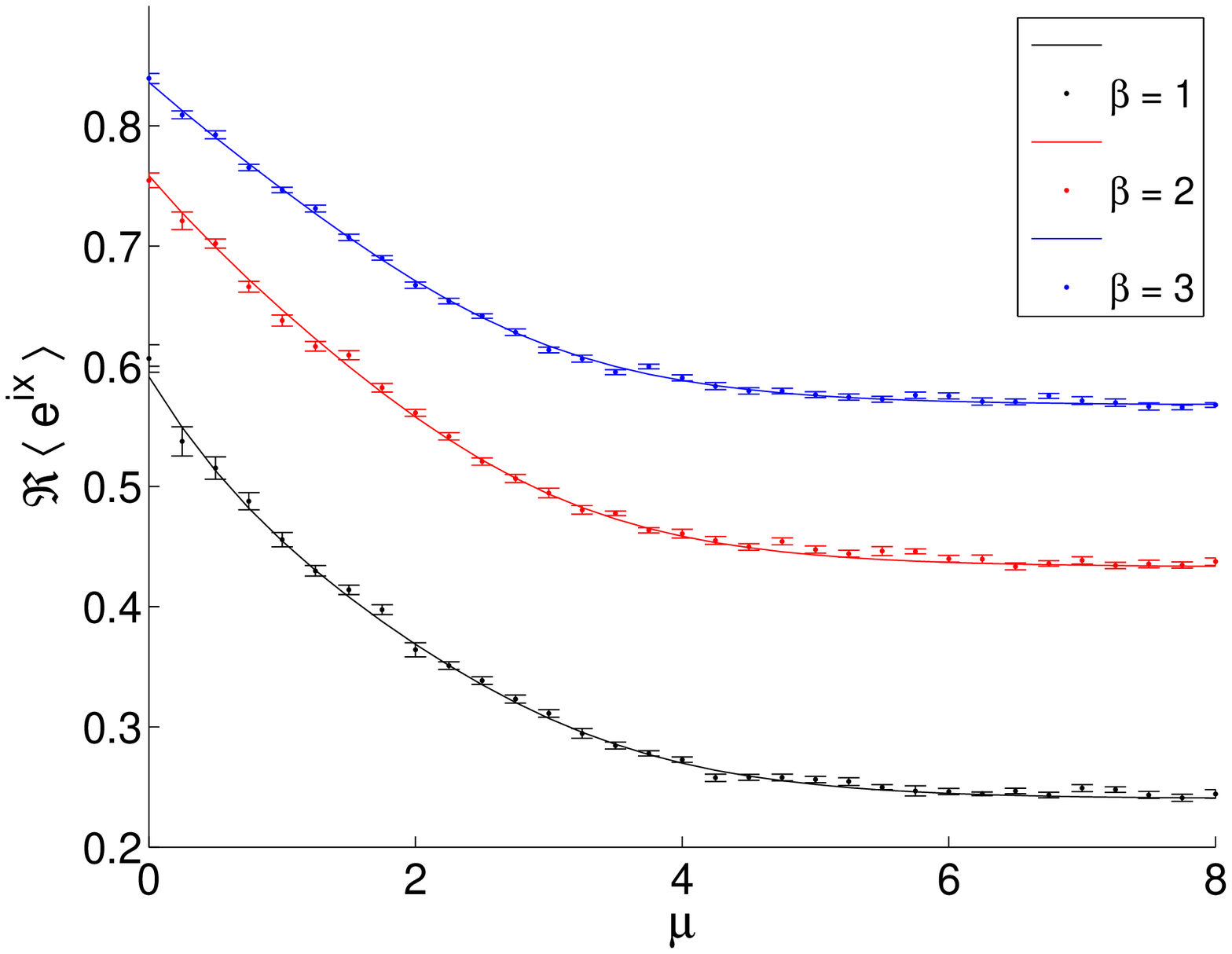}
}
\subfloat[][The conjugate Polyakov loop $\langle e^{-ix} \rangle$]{
	\includegraphics[scale=0.23]{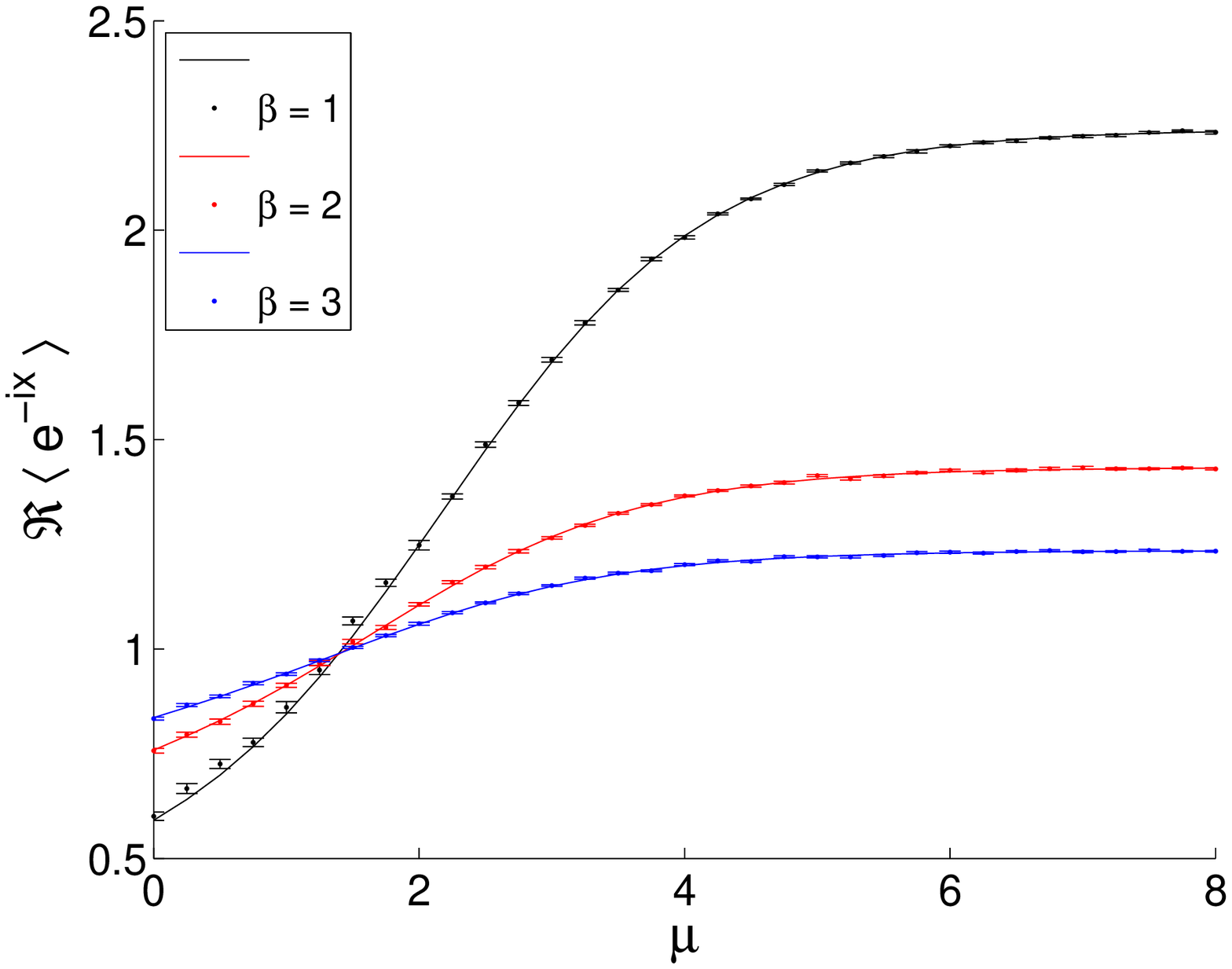}
} \\
\subfloat[][The density $\langle n \rangle$]{
	\includegraphics[scale=0.23]{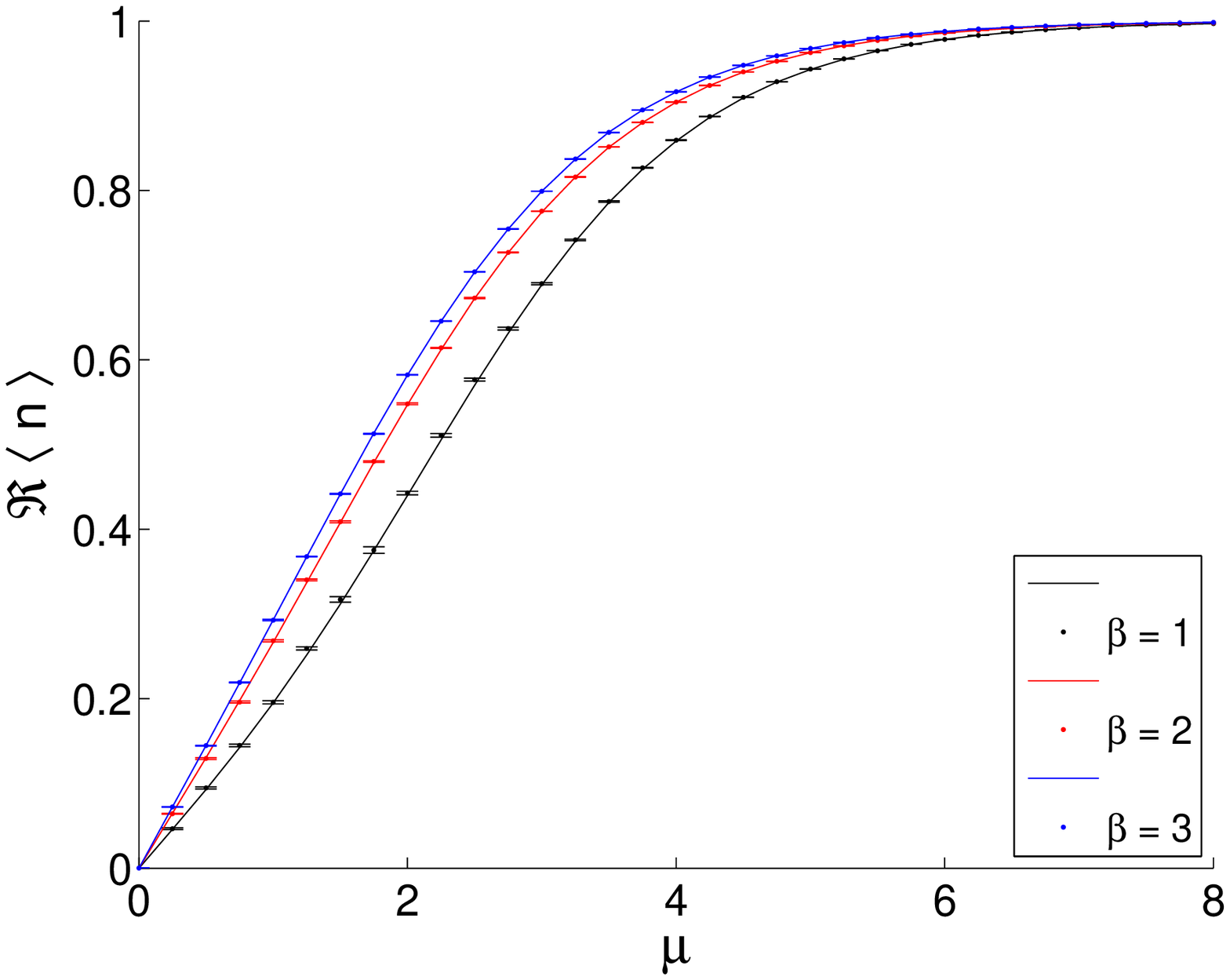}
}
\subfloat[][The plaquette $\langle \cos \left( x\right) \rangle$. Real Langevin was used to estimate the observable for imaginary $\mu$, hence negative $\mu^2$.]{
	\includegraphics[scale=0.23]{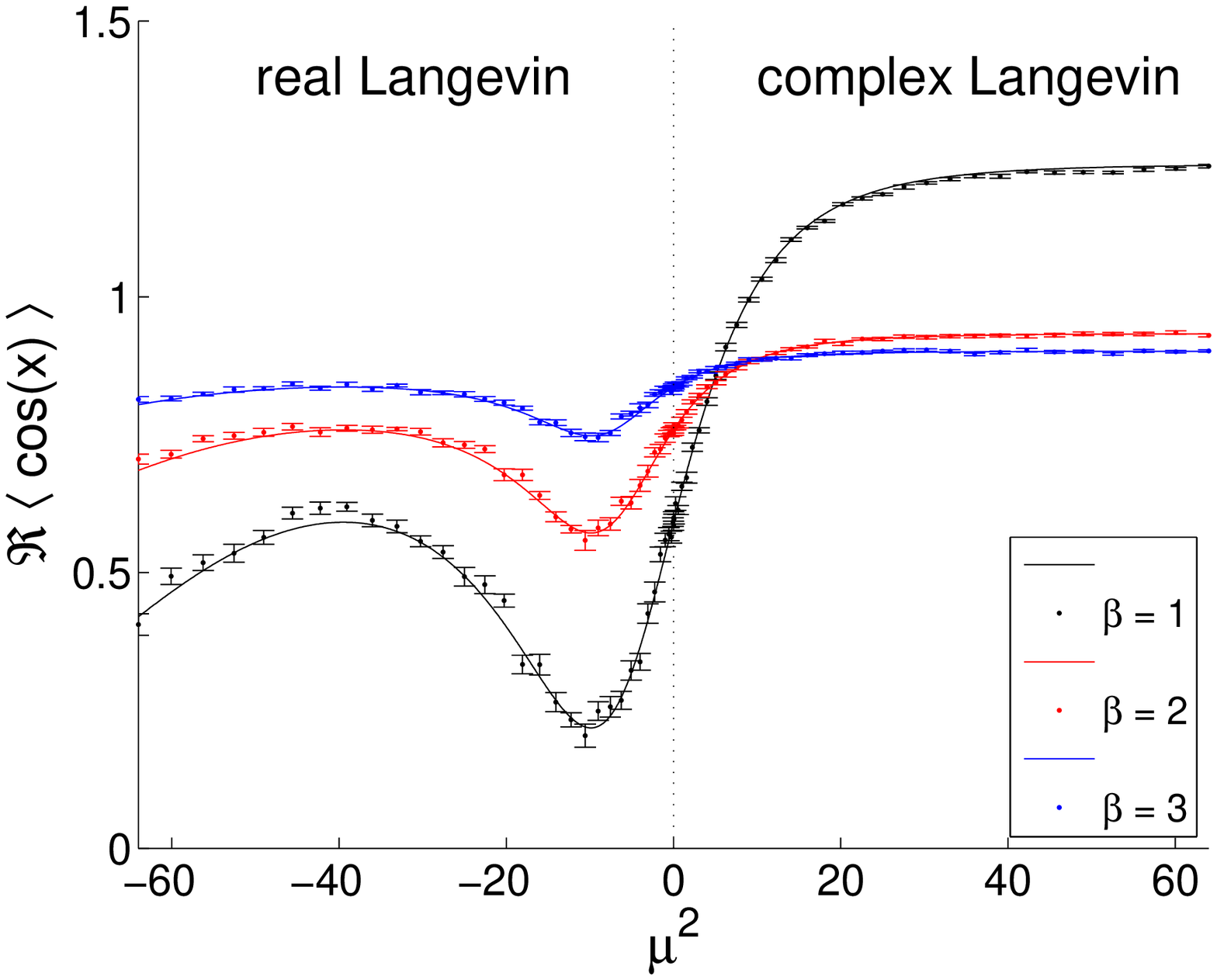}
	\label{fig:plaquette}
}
\caption{Estimates of observables vs. $\mu$ for parameters $\kappa = 0.5$, $\beta= 1$, $2$ and $3$, $\epsilon = 5\cdot 10^{-5}$ and for an average of $20$ trajectories, each $N = 2.5 \cdot 10^6$ time steps long. Here, the success of complex Langevin dynamics is confirmed.}
\label{fig:orgobs}
\end{figure}

In accordance with \cite{aartsu1}, the observables
\begin{align}
\langle e^{ix} \rangle &= \frac{I_1(\beta) + \kappa I_1'(\beta) \cosh(\mu) - \kappa I_1(\beta) \beta^{-1} \sinh(\mu)}{I_0(\beta)+\kappa I_1(\beta) \cosh(\mu)}\textrm{,} \\
\langle e^{-ix} \rangle &= \langle e^{ix} \rangle |_{\mu \to -\mu}\textrm{,} \\
\langle \cos(x) \rangle &= \frac{1}{Z} (I_1(\beta) + \kappa I_1'(\beta) \cosh(\mu))\textrm{,} \\
\langle n \rangle &= \frac{1}{Z} \kappa I_1(\beta) \sinh(\mu)\textrm{,}
\end{align}
where $Z = I_0(\beta)+\kappa I_1(\beta)\cosh(\mu)$ and $I_n$ are the modified Bessel functions of the first kind of order $n$, are also here found to be estimated correctly using complex Langevin dynamics for $\kappa = 0.5$. The results are shown in Figure~\ref{fig:orgobs} for a good range of $\mu$ and $\beta$, and the agreement is convincing. In Figure~\ref{fig:plaquette}, the plaquette is shown as a function of the squared chemical potential which when negative corresponds to a real action and makes real Langevin dynamics possible. We see that the transition about $\mu = 0$ is smooth.

\section{Limitations of complex Langevin dynamics}
\label{sec:limits}
\begin{figure}
\centering
\subfloat[][Complex Langevin dynamics]{
	\includegraphics[scale=0.23]{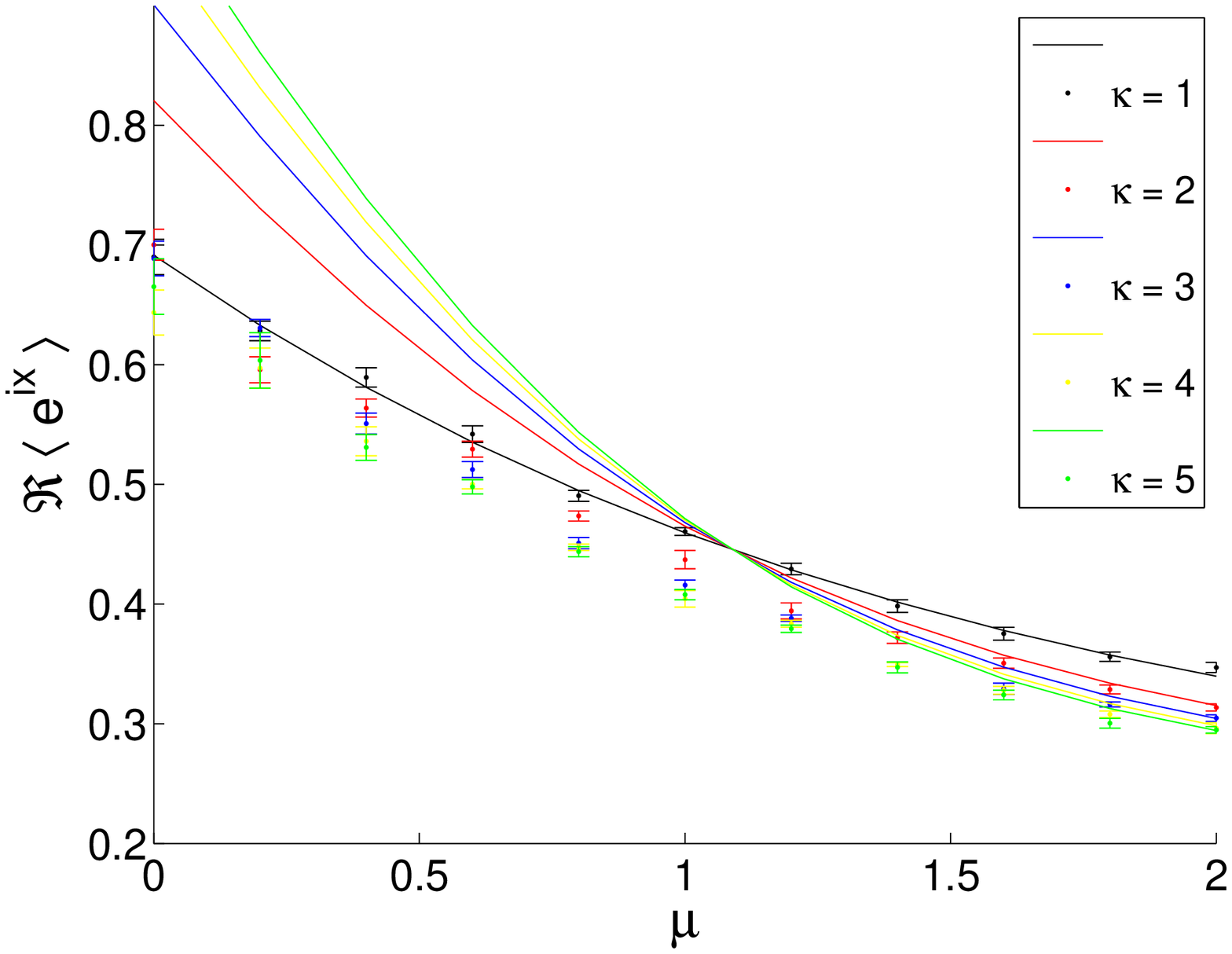}
	\label{fig:cld}
} 
\subfloat[][Real Langevin dynamics with reweighting]{
	\includegraphics[scale=0.23]{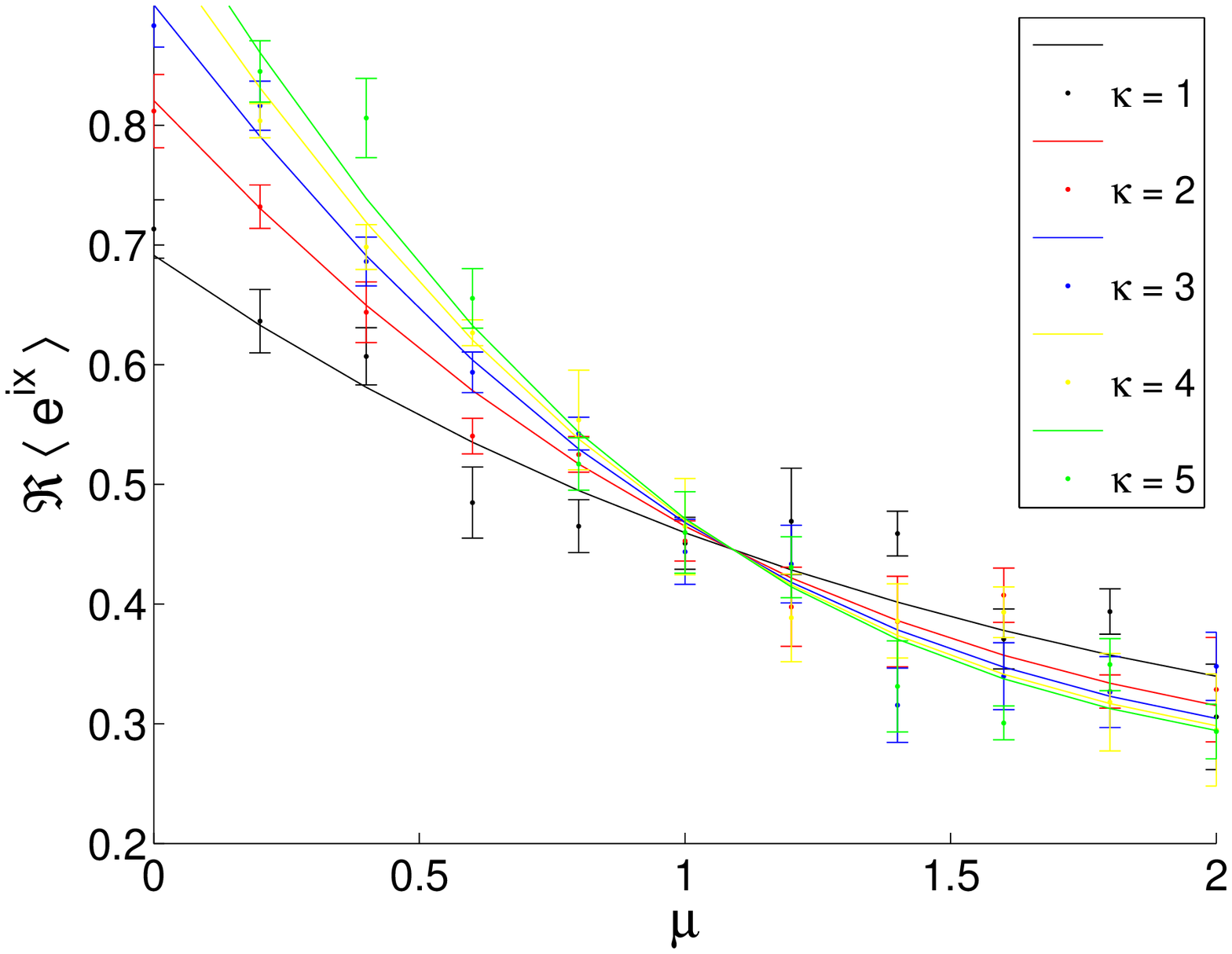}
	\label{fig:rldr}
}
\caption{Estimates (dots) of $\langle e^{ix} \rangle$ vs. $\mu$ for different values of $\kappa$ compared to their analytical values (full lines). Each colour corresponds to a different value of $\kappa$ and $\beta$ was in all cases set to $1$.}
\label{fig:kapvar}
\end{figure}

\begin{figure}[H]
								\centering
								\subfloat[][The first power $\langle x \rangle$]{
								\includegraphics[scale=0.23]{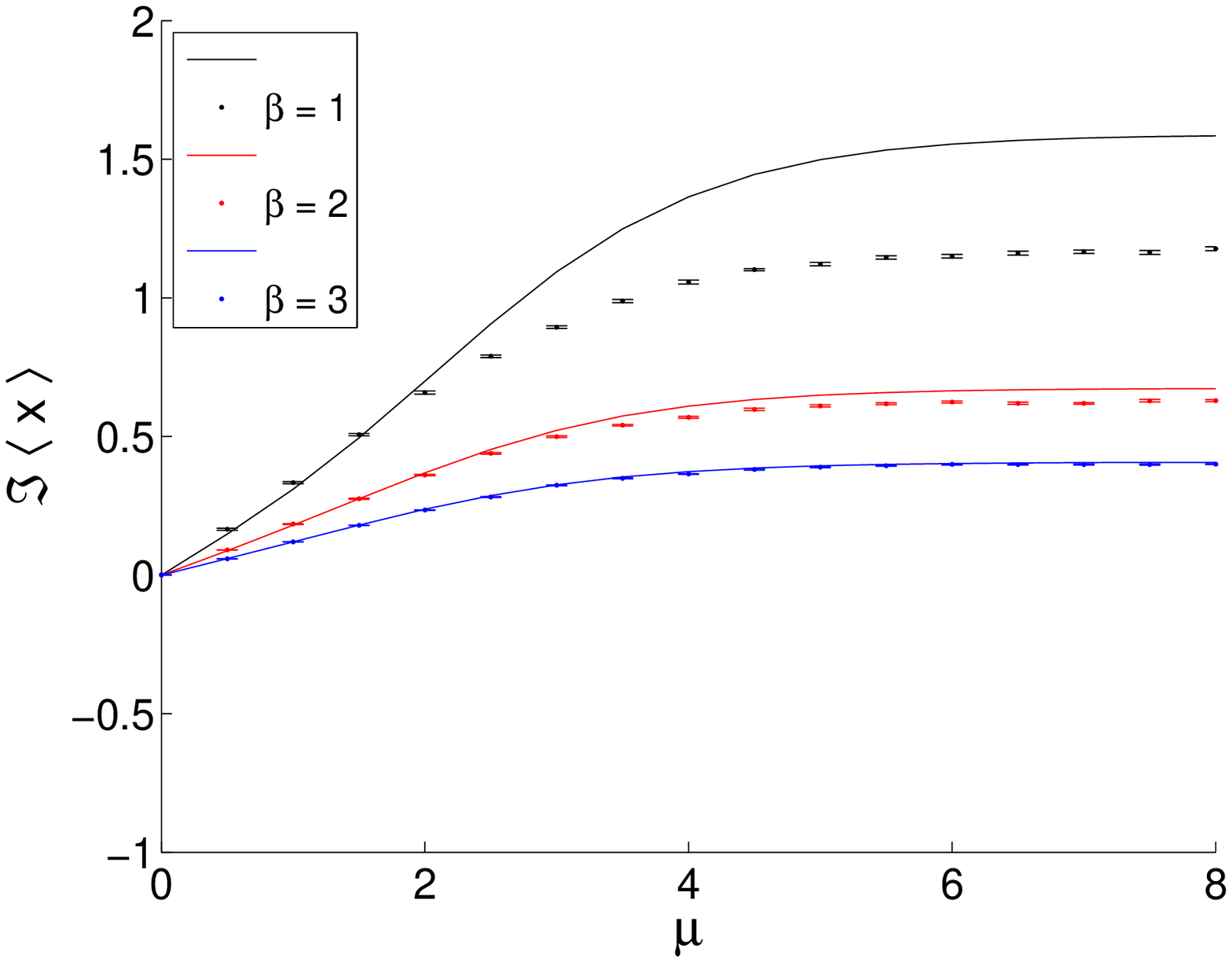}
								\label{fig:x1}
								}
								\subfloat[][The second power $\langle x^{2} \rangle$]{
								\includegraphics[scale=0.23]{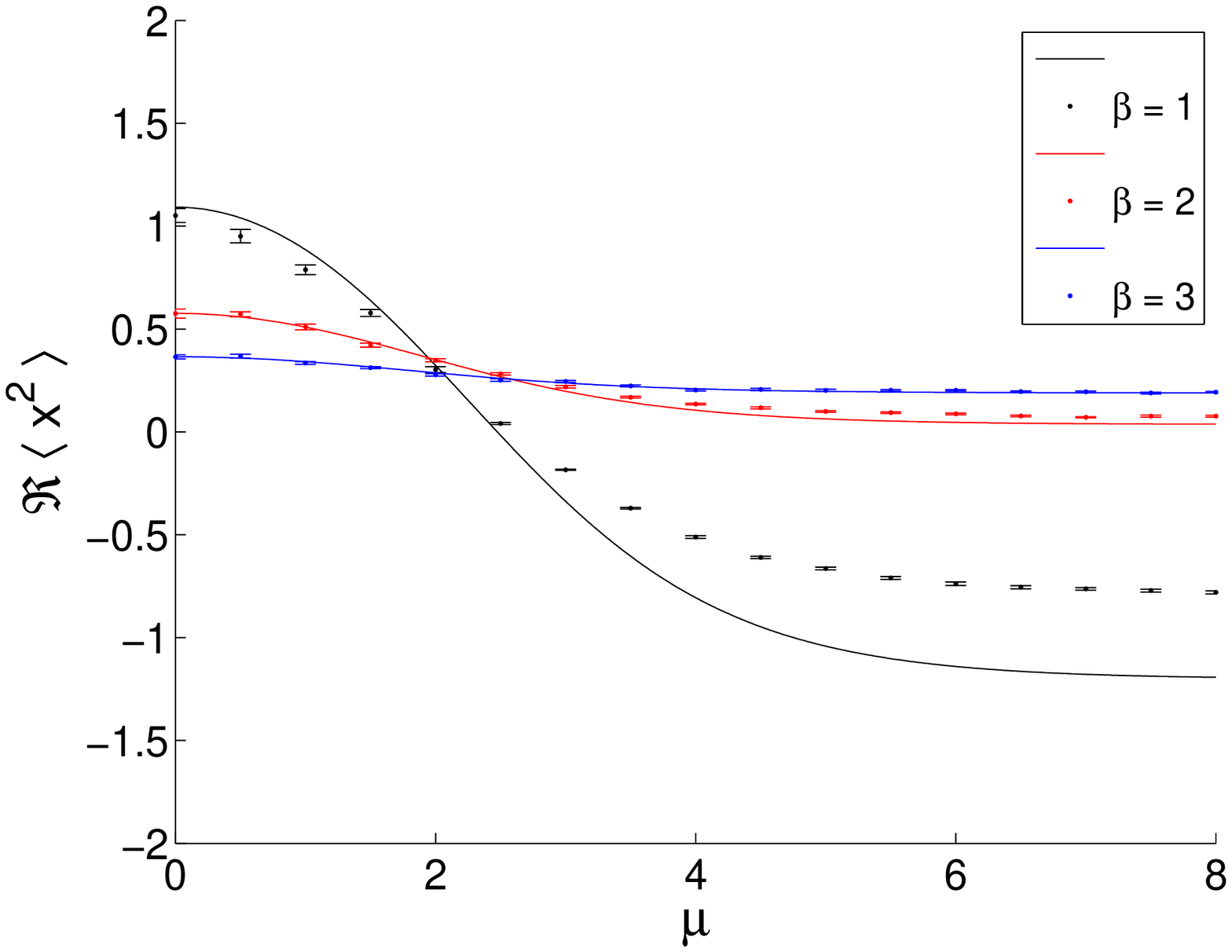}
								\label{fig:x2}
								} \\
								\subfloat[][The third power $\langle x^{3} \rangle$]{
								\includegraphics[scale=0.23]{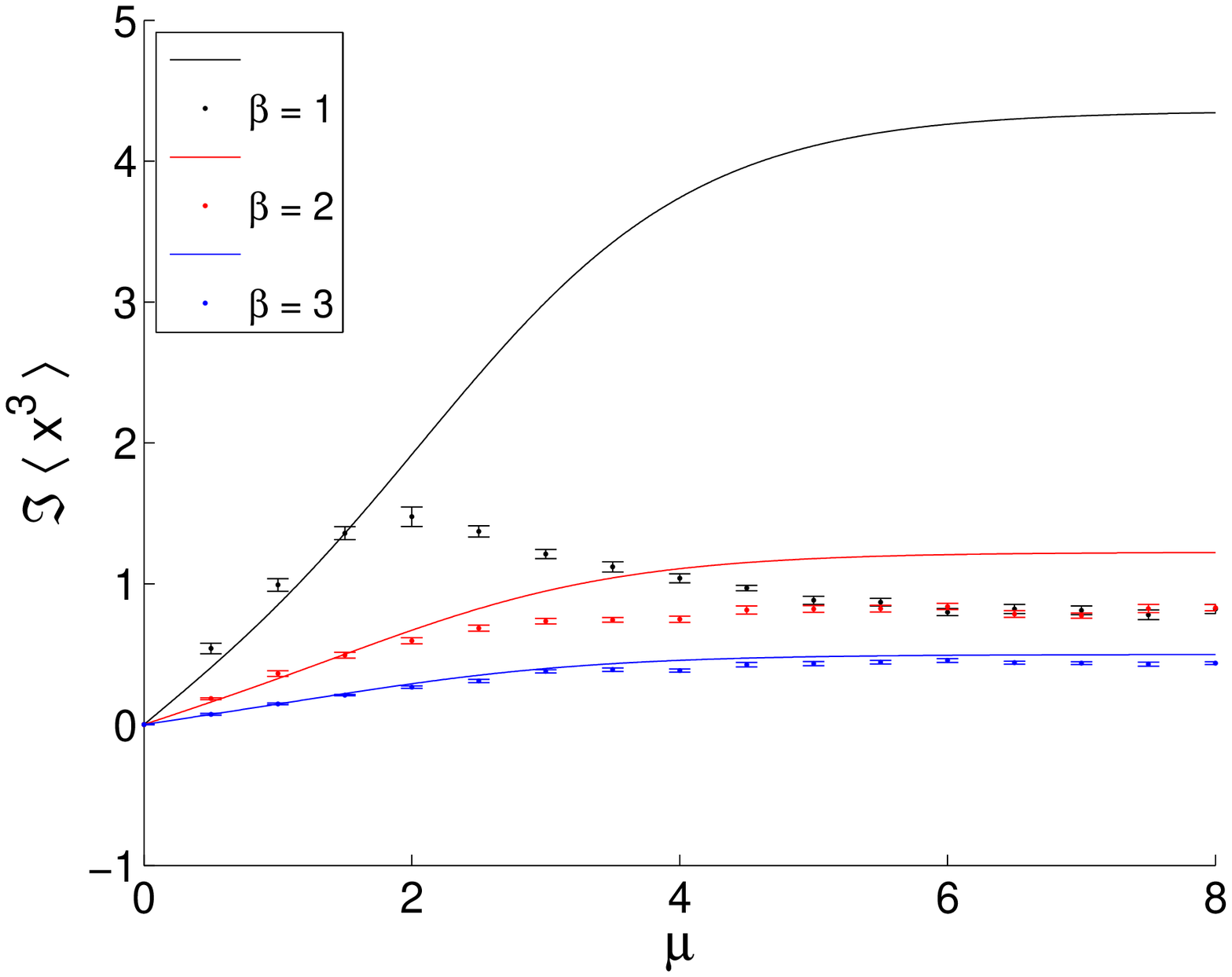}
								\label{fig:x3}
								}
								\subfloat[][The fourth power $\langle x^{4} \rangle$]{
								\includegraphics[scale=0.23]{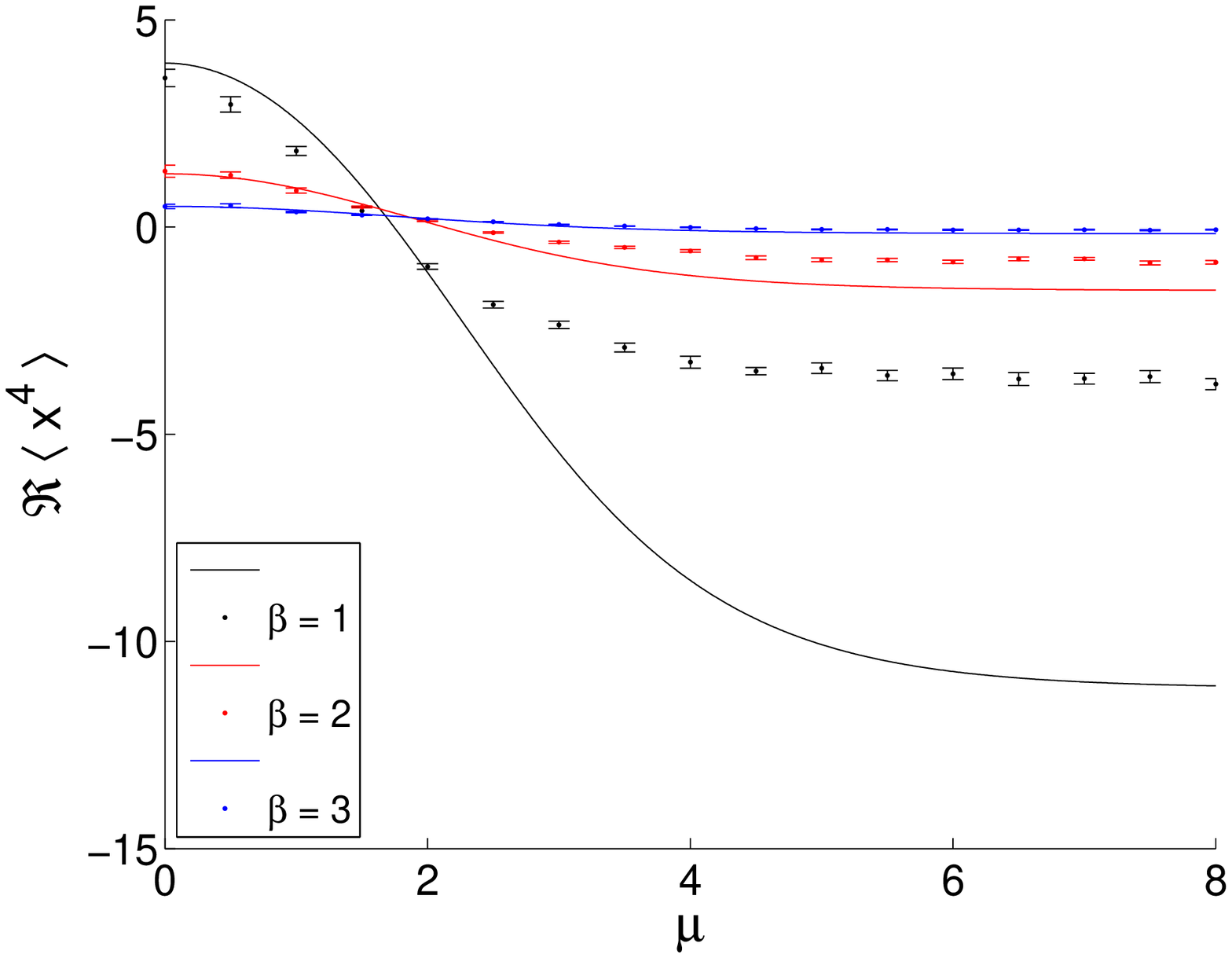}
								\label{fig:x4}
								}
								\caption{Estimates (dots) of the first four moments using complex Langevin dynamics compared to expected values obtained by numerical integration (full lines).}
								\label{fig:x14}
							\end{figure}

\begin{figure}[H]
								\centering
								\subfloat[][Complex Langevin dynamics]{
								\includegraphics[scale=0.23]{1mean.eps}
								\label{fig:cmean}
								}
								\subfloat[][Real Langevin dynamics with reweighting]{
								\includegraphics[scale=0.23]{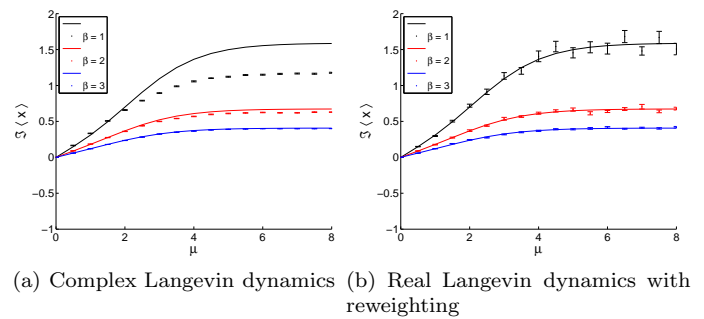}
								\label{fig:rmean}
								}
								\caption{Estimates (dots) of $\langle x \rangle$ vs. $\mu$ for different $\beta$ as compared to the exact values (full lines). In \protect\subref{fig:cmean} the estimates were obtained by complex Langevin dynamics and in \protect\subref{fig:rmean} by reweighting real Langevin dynamics.}
								\label{fig:estx}
							\end{figure}

\begin{figure}
								\centering
								\subfloat[][Third order estimate of $\langle e^{ix} \rangle$]{
								\includegraphics[scale=0.23]{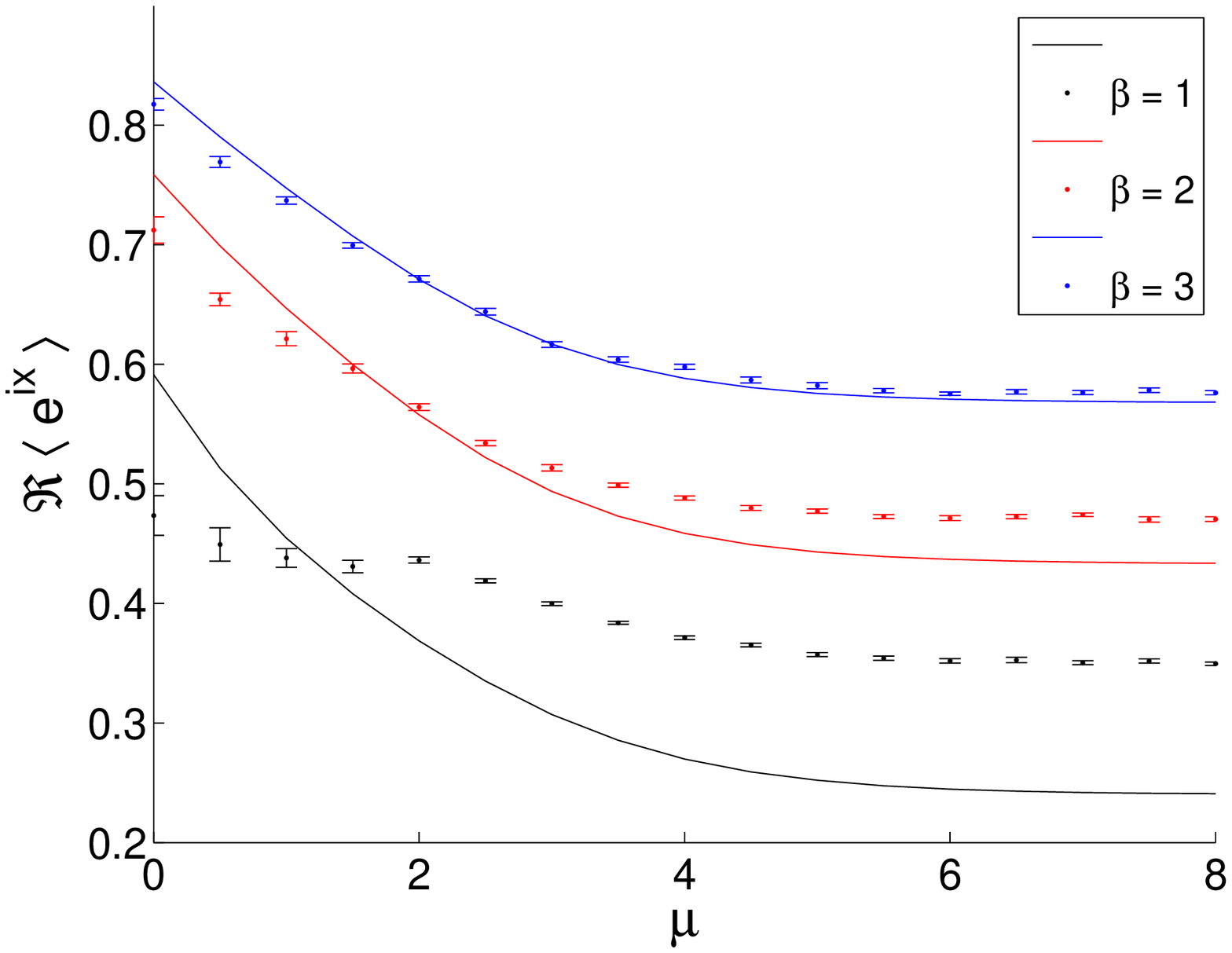}
								\label{fig:xpon3}
								}
								\subfloat[][Seventh order estimate of $\langle e^{ix} \rangle$]{
								\includegraphics[scale=0.23]{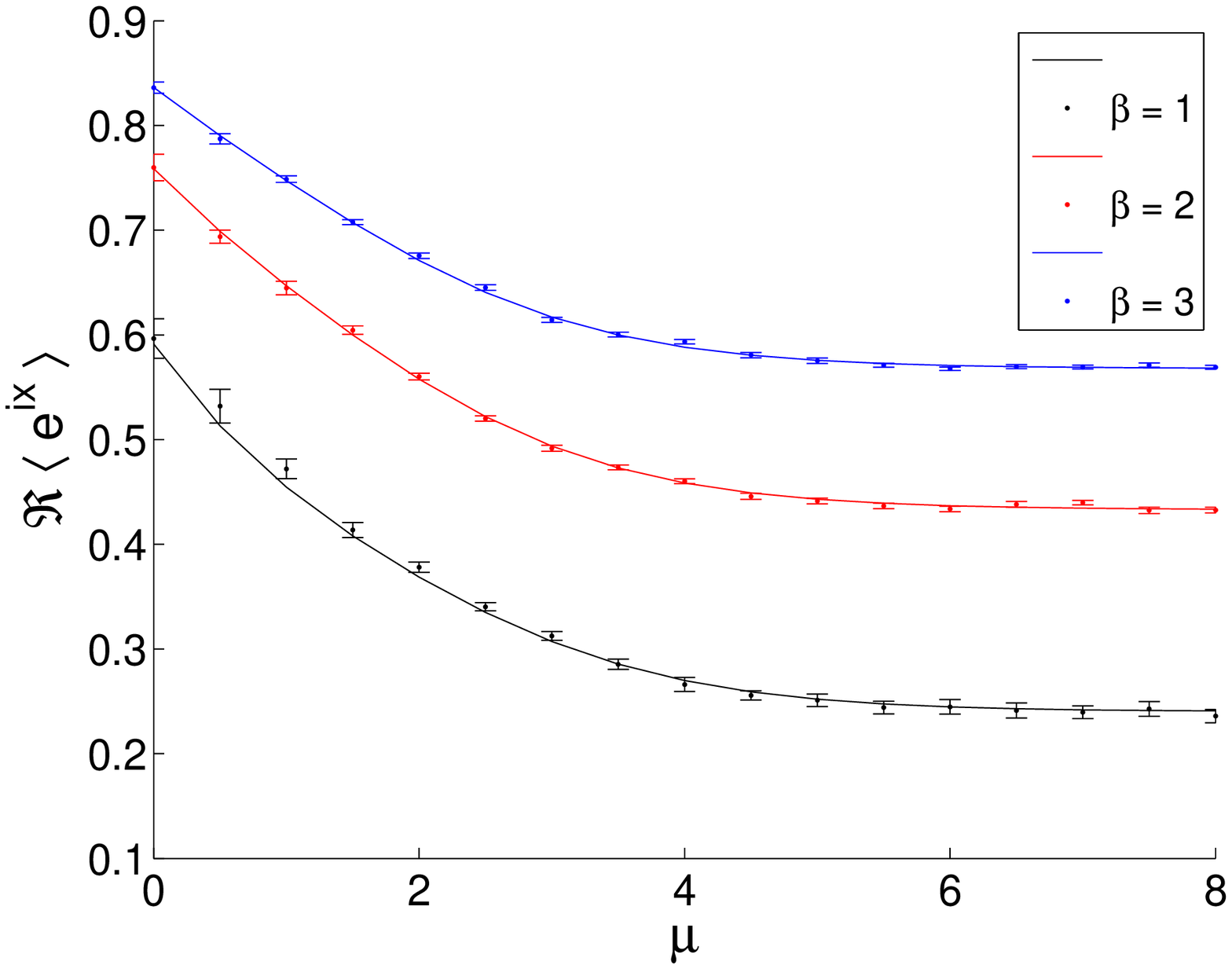}
								\label{fig:xpon7}
								}
								\caption{Estimates (dots) of $\langle e^{ix} \rangle$ vs. $\mu$ for different $\beta$ as compared to the exact values (full lines). In \protect\subref{fig:xpon3} the first three moments were used to form the estimate and in \protect\subref{fig:xpon7} the first seven moments were used.}
								\label{fig:estxpon}
	\end{figure}

\begin{figure}[H]
								\centering
								\subfloat[][Real part of $\langle e^{x} \rangle$]{
								\includegraphics[scale=0.23]{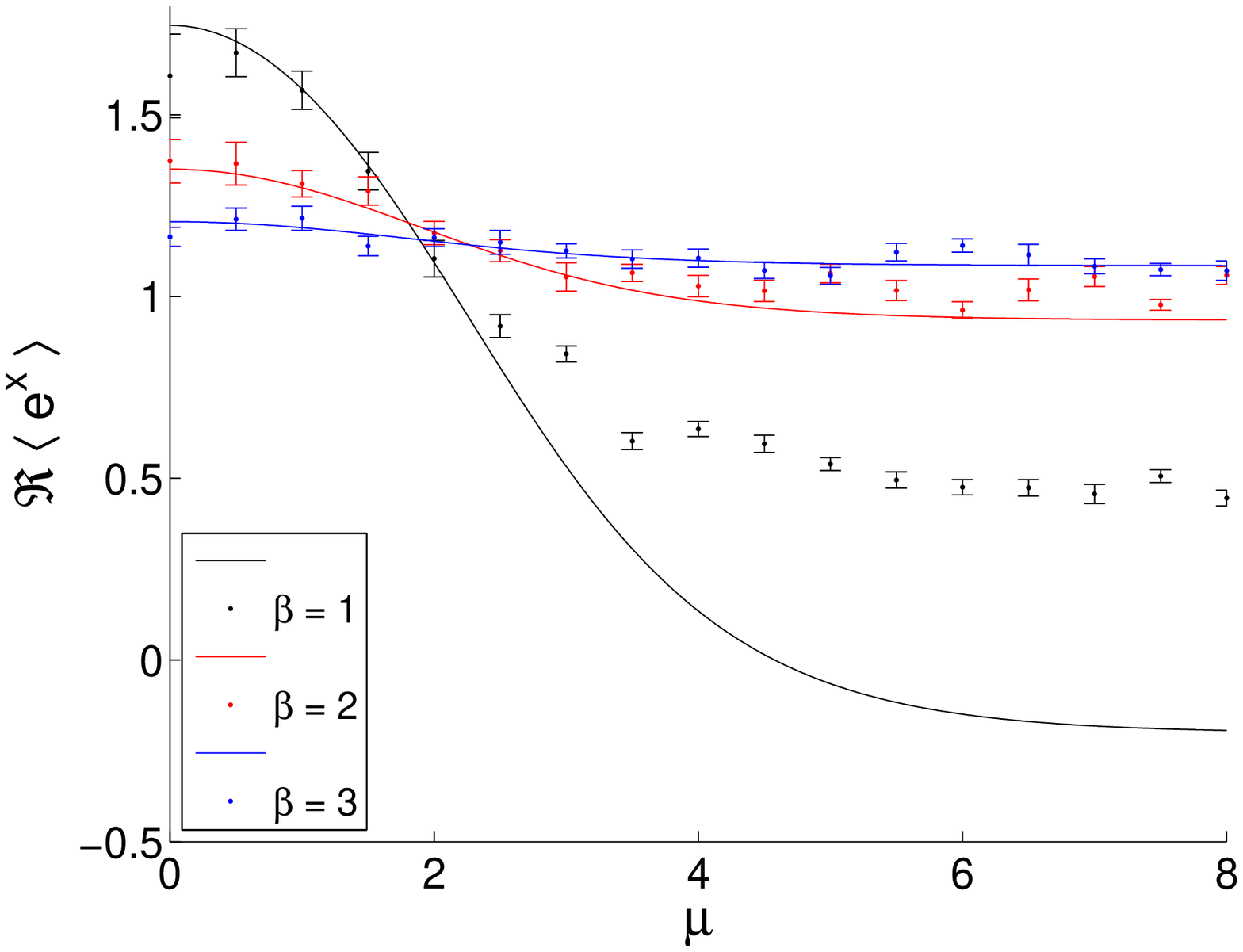}
								\label{fig:expre}
								}
								\subfloat[][Imaginary part of $\langle e^{x} \rangle$]{
								\includegraphics[scale=0.23]{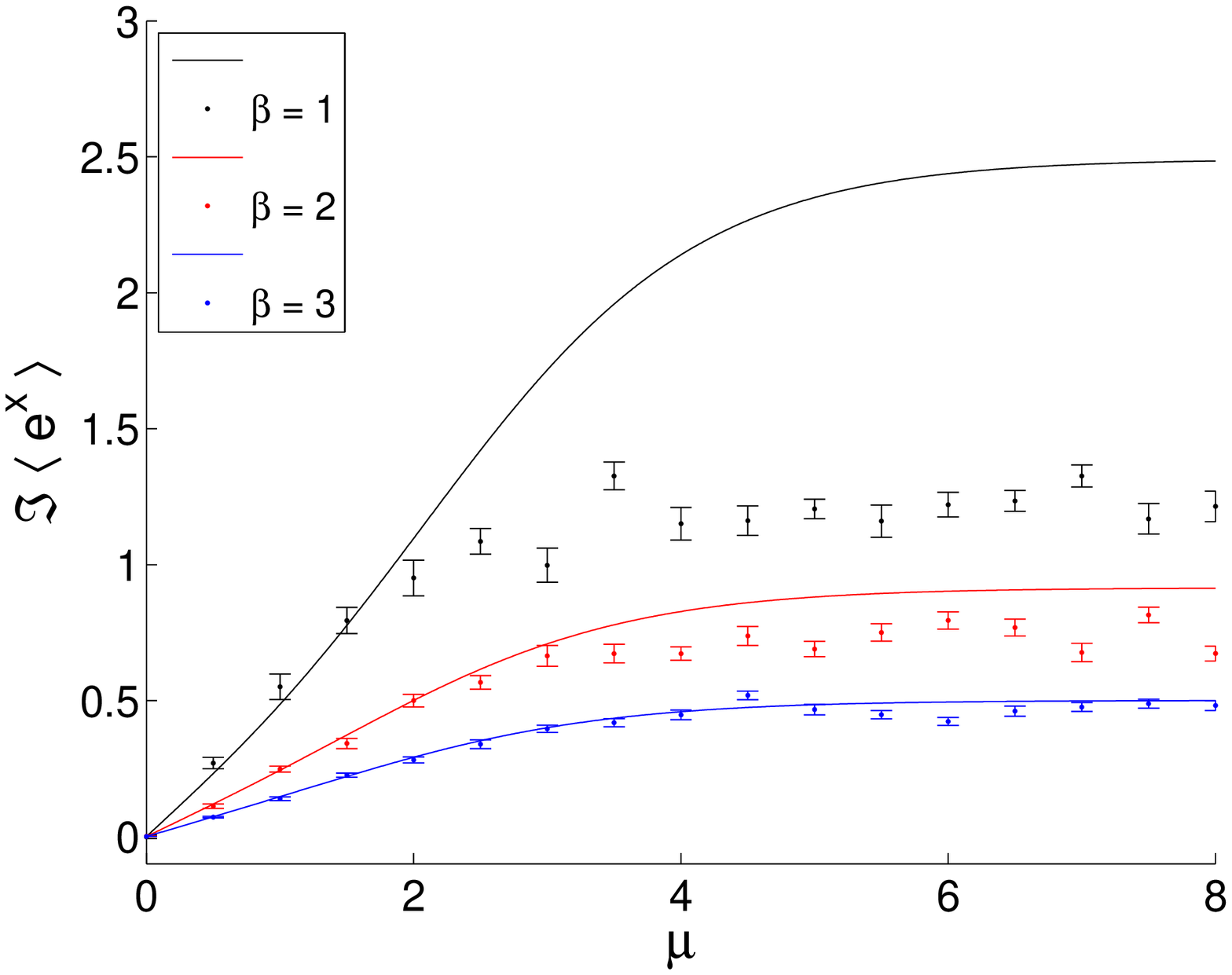}
								\label{fig:expim}
								}
								\caption{Estimates (dots) of $\mathrm{Re}\, \langle e^{x} \rangle$ and $\mathrm{Im} \, \langle e^{x} \rangle$ vs. $\mu$ for different $\beta$ as compared to the expected values obtained by numerical integration (full lines).}
								\label{fig:reimexp}
	\end{figure}

It was noted in \cite{kim} that complex Langevin dynamics fails to correctly estimate $\langle e^{ix} \rangle$ as $\kappa$ is increased from the previously studied value of $\kappa = 0.5$. The discrepancies are particularly significant for values of $\mu$ in the interval $[0,2]$ but convergence is again restored as $\mu$ is increased. The results are shown in Figure~\ref{fig:kapvar} and it is suggested in \cite{kim} that it is due to the nonanalyticity of the logarithm at the origin $(x,y)=(0,0)$. For small $\mu$ and/or large $\kappa$, the Langevin trajectory probes the region near the origin whereas for large $\mu$ and/or small $\kappa$ it does not. The results obtained using complex Langevin dynamics are shown in Figure~\ref{fig:cld} where there is almost no distinction between different values of $\kappa$, and complex Langevin dynamics clearly fails to reproduce the analytic curves (full lines). In Figure~\ref{fig:rldr}, we see the same observable obtained using real Langevin dynamics and reweighting. The agreement is much better albeit with a larger uncertainty.
Conversely, the argument of \cite{kim} suggests that as long as the Langevin flow is not in the neighbourhood of the origin, complex Langevin dynamics should work well. As we will see, this is not necessarily true. Consider the simple moments $\langle x^n \rangle$. For $n=1$, there exists an analytic expression equal to
\begin{align}
\langle x \rangle = \frac{i e^{-\beta} \kappa (-1 + e^{\beta} I_0(\beta)) \sinh(\mu)}{\beta (I_0(\beta)+ \kappa I_1(\beta)\cosh(\mu))}
\end{align}
but simple numerical integration schemes may be used to evaluate higher powers.
We computed these using complex Langevin dynamics (see Figure~\ref{fig:x14}), taking $\kappa = 0.5$ for which we found the correct results of Figure~\ref{fig:orgobs}.
Since the functions $x^n$ are not periodic but the model considers only values of $x$ within $[-\pi,\pi]$ in the original expression for $\langle O \rangle$, instead, a periodic extension of $x^n$ was considered, just replacing all other intervals $[-\pi+2\pi m,\pi+2\pi m)$ where $m \in \mathbb{Z}$ with the definition of $x^n$ in the interval $[-\pi,\pi)$. Observables were complexified $x^n \to (x+iy)^n$ and averaged over the trajectory in the complex plane defined by complex Langevin dynamics. Expectation values of even powers are on average real whilst those of odd powers are imaginary. This follows from considering $(x+iy)^n$ which has terms $i^k x^{n-k} y^k$. There are only even contributions with respect to $x$ because of the symmetry of the action in $x$. This means that if $n$ is even, and only even $n-k$ contribute, then $k$ must be even and hence $i^k$ is real. The results are shown in Figure~\ref{fig:x14}.
The discrepancies are most visible when $\beta = 1$. The errors could not be reduced by increasing the finite run time. The trend seems to be general. There is an initial small overestimation (underestimation when $n$ is even), followed by an underestimation (overestimation when even) that becomes significant with an increase in $\mu$ but seems to tend to a constant. This is also observed for higher powers $x^5$, $x^6$ and $x^7$ but it is conjectured to hold for $n$.
This discrepancy is only present for complex Langevin dynamics. There is no difficulty estimating these powers using real Langevin dynamics with reweighting. This is clear from Figure~\ref{fig:estx} where the estimates of $\langle x \rangle$ due to complex Langevin dynamics and real Langevin dynamics with reweighting are compared.
These results could just be accepted and a concession could be made that complex Langevin dynamics defined on compact spaces simply fails to properly estimate observables that are not \emph{ab initio} periodic and that estimates of powers $\langle x^n \rangle$ are useless. However, it is curious that these misestimates combined can be used to produce a result that is correct. These misestimated powers may be used to reconstruct the successful $\langle e^{ix} \rangle$ by its Taylor series $1+ i \langle x \rangle - \frac{1}{2} \langle x^2 \rangle + \ldots$ and the results are shown in Figure~\ref{fig:estxpon} where $\langle e^{ix}\rangle$ is reconstructed by including terms up to third order and seventh order.
Though $\langle e^{\pm ix} \rangle$ is estimated correctly using complex Langevin dynamics, $\langle e^{\pm x}\rangle$ is not. In light of the observations that odd powers are, as a function of $\mu$, initially overestimated and subsequently underestimated, and vice versa for even powers, it is clear that $\langle e^{\pm x}\rangle$ should be misestimated. Without the imaginary element included in the exponent of $\langle e^{\pm x} \rangle$, overestimates may never cancel underestimates. When the imaginary element is included, $\langle i^{2k+1} x^{2k+1} \rangle $ which is real and initially overestimated may be paired with terms such as $\langle i^{2k} x^{2k}\rangle$ which is also real but initially underestimated and a correct convergence is possible. This is not so with $\langle e^{\pm x} \rangle$ which segregates initially overestimated and underestimated parts into real and imaginary parts, respectively. Indeed, the observables $\langle e^{\pm x} \rangle$ are also misetimated using complex Langevin dynamics. The results of $\mathrm{Re} \, \langle e^{x} \rangle$ and $\mathrm{Im} \, \langle e^{x} \rangle$ are both nonvanishing and shown in Figure~\ref{fig:reimexp}.
$\langle e^{x} \rangle$ is related to $\langle e^{-x} \rangle$ by an overall sign in the imaginary part. Another possible way to have misestimates cancel one another is by an alternating series in the odd or even moments such as happens for $\langle \sin(x) \rangle$ and $\langle \cos(x)\rangle$.

\section{Conclusion}
\label{sec:concl}
Complex Langevin dynamics is a promising generalisation of the well-established stochastic quantisation procedure to complex-valued actions. But although successful in a number of cases, even for simple one-dimensional models as the one considered here, we have seen that there are unsettled issues. Attracting most attention has been the issue of convergence of the Langevin process in order to generate the correct probability distribution. As described in \cite{kim}, one is the ambiguity of the logarithm in the action which does not carry over to the equations of motion and may therefore produce wrong results. There may also be runaway solutions, or a well-behaved distribution may be obtained, which is however the wrong one \cite{BergesMinkowski}. 

As we describe here, only certain observables give the right results on an otherwise unproblematic Langevin trajectory not affected by the logarithmic ambiguity. But remarkably, even when certain observables evaluate to the wrong values (in our case the moments $x^n$), they are wrong in such a way that they combine as a Taylor series into certain other correctly reproduced observables. This is reminiscent of distinguishing between gauge invariant and gauge-noninvariant observables, although we are not convinced that the analogy carries through completely. It shows that the probability distribution $P(x,y)$ does have information to correctly reproduce some observables, but not others, so $P(x,y)$ is not a complete representation of the complex path integral computation in the same way that the reweighting method is. In particular, $P(x,y)$ cannot be reproduced from the complex-valued action by matching moment by moment. The relationship is more subtle and still unclear, and we believe it merits further investigation.

\bibliographystyle{elsarticle-num}
\bibliography{references}
\end{document}